# HVDC Surface Flashover in Compressed Air for Various Dielectrics

**I. A. Bean** and **C. S. Adams**
Virginia Polytechnic Institute and State University
Virginia, US

**T. E. Weber**
Los Alamos National Laboratory
New Mexico, US

## ABSTRACT

This study measures the voltage at which flashover occurs in compressed air for a variety of dielectric materials and lengths in a uniform field for DC voltages up to 100 kV. Statistical time lag is recorded and characterized, displaying a roughly exponential dependence on breakdown voltage. Of the materials tested, acrylic is observed to be the most resistant to flashover. These data are intended to facilitate the design of compressed-air insulated high voltage systems as an alternative to $SF_6$ insulated systems.

## 1 INTRODUCTION

SURFACE flashover is a failure mode that occurs in high voltage equipment such as gas-insulated transmission lines and spark gap switching technology. Past experimental studies of flashover conducted in a wide range of settings [1 - 4] have found that the initiation of surface flashover is dependent on a number of factors including electrode material and surface condition, spacer material and surface condition, fluid composition and pressure, waveform of the applied potential, and the geometry of the electrode/spacer setup.

Sulfur hexafluoride ($SF_6$) is commonly used as an insulating gas due to its high dielectric strength. However, $SF_6$ is a potent greenhouse gas with a global warming potential ranging from 16,300 to 32,600 times that of $CO_2$ for time horizon estimates ranging from 20 to 500 years [5]. Regulations regarding the use of $SF_6$ are becoming increasingly stringent [6], which incentivizes exploration of environmentally acceptable alternatives. Previous work in this area [7] has surmised that compressed air is viable for electrical insulation; however, its use would require testing and design of new systems to replace present $SF_6$ systems. $SF_6$ has been well-studied with respect to surface flashover [8 - 11] and, while past research has covered flashover in air at pressures ranging from one atmosphere down to vacuum for insulated systems [12, 13], there exists a lack of published data on flashover characteristics of compressed-air insulated systems. Furthermore, many DC flashover studies do not record the statistical time lag, the elapsed time prior to flashover after the application of sufficient voltage, which is crucial to the system design since the statistical time lag to flashover for a system should be orders of magnitude greater than the time-span for which the system will remain in a charged state.

The experiments described in this paper investigate surface flashover between electrodes separated by an insulating spacer in air up to pressures of 490 kPa (71 psia) in a uniform field distribution standing off DC potentials up to 100 kV with varying spacer materials and lengths. The statistical time lag (the average time required for breakdown to occur after the application of voltage) is also recorded and characterized for timescales of 30 seconds or less, a time frame relevant to charging pulsed-power systems. The parameter regime investigated is applicable to conditions in spark gap switch technology and results contribute to the present efforts at Los Alamos National Laboratory to design a new pulsed-power infrastructure employing compressed-air-insulated field-distortion switches.

The remainder of this paper is organized as follows: Section 2 discusses basic theory and experimental results of previous work to provide context for the experimental design; Section 3 details the configuration of the experiment and the methodology used; Section 4 presents the results and discusses their implications; finally, Section 5 discusses conclusions and outstanding questions.

## 2 BACKGROUND

Surface flashover refers to breakdown of a gas across the surface of an insulator under an applied electric field. Gas-insulated high-voltage systems require spacers to separate electrodes and these spacers introduce a surface over which flashover can occur. The presence of a spacer tends to reduce the voltage at which breakdown will occur when compared to the theoretical breakdown voltage in the absence of a spacer.

The theoretical breakdown voltage of a gas between two electrodes in the absence of a spacer can be determined using

Paschen's law, which relates the product of pressure and distance to the voltage at which the gas will ionize and conduct (assuming a uniform electric field distribution). Paschen's law predicts the breakdown voltage in a gas with a uniform electric field to be:

$$V_b = \frac{Bpd}{\ln(pd) + k}, \quad (1)$$

where

$$k = \ln\left(\frac{A}{\ln\left(1 + \frac{1}{\gamma}\right)}\right), \quad (2)$$

and where $p$ is the pressure in kPa, $d$ is the distance in cm, $A$ is the gas saturation ionization constant in (kPa·cm), $B$ is the gas ionization energy constant in V/(kPa·cm), and $\gamma$ is the secondary emission coefficient of the electrode. This paper compares experimental measurements of flashover to Paschen's Law using the $A$ and $B$ coefficients obtained by Husain [14] for air where $A = 112.5$(kPa·cm)$^{-1}$ and $B = 2737.5$ (V/(kPa·cm)). The secondary emission coefficient for copper in air (an approximation to brass) was obtained from Cobine [15] as $\gamma \approx 0.025$.

Surface charging of the spacer is thought to play a dominant role in flashover [2] and reduces the effective gap distance by a length defined as the "analogous ineffective region" by Li [16]. Surface charge accumulation can be caused by charge migration through volumetric and surface conductivity as well as free electrons impacting, and subsequently becoming trapped along the spacer surface [17]. However, conductive charge migration typically contributes to surface charge accumulation over much longer timescales when compared to free electron sources [18]. Therefore, for the 30 second timescale explored in this study, free electron sources are expected to dominate surface charge accumulation, but conductive charge migration should not be discounted. Free electron sources can include:

- natural ionization due to background radiation
- secondary emission from electrode and spacer surfaces [19]
- field emission from electrodes [20]
- partial discharges in the gas [4, 21, 22]

However, field emission and partial discharges are expected to dominate other sources of free electrons at the pressures investigated here due to short mean-free-paths, which limit the ion impact energy available to induce secondary emission.

Although background radiation can provide seed electrons for partial discharges and ionization, it is not the dominant source of free electrons leading to surface charging. Secondary emission of electrodes is accounted for by $\gamma$ in the Paschen law and does not significantly affect breakdown voltage calculations at higher pressures due to shorter mean-free-paths. Secondary emission from the spacer is also mitigated by shorter mean-free-paths at higher pressures. It has been shown that field enhancement near the triple-junction (the location of the electrode, insulator, and gas interface) can yield electric field intensities capable of initiating field emission [11]. This field enhancement can result from imperfect mating of electrode and insulator surfaces as well as microscopic electrode surface protrusions, i.e., surface condition. The degree of enhancement is also dependent on both the work function of the electrode material and the permittivity of the spacer material [17, 20, 23]. Partial discharges in the gas result from the initiation of a Townsend avalanche [15] that does not generate sufficient electron density to form and sustain an arc across the gap. Partial discharges that occur near the spacer surface can provide a significant source of free electrons for surface charging at higher pressures. It should be noted that surface roughness of the spacer causes local field enhancement that scales with its permittivity and can encourage the formation of partial discharges. Therefore, the most probable sources of free electrons for surface charging in the present study include field emission at the triple-junction and partial discharges along the spacer surface, both of which scale with material permittivity. However, the relative timescales on which surface charging occurs for varying pressures, gas compositions, and dielectric materials are not yet well understood.

Charge migration through bulk and surface conductivity can also influence surface charging as shown by Li [24]. The presence of normal electric fields along the spacer's surface can encourage charge migration from the spacer interior to the surface, yielding another source of surface charging. It has also been shown that normal fields lead to higher dispersion of flashover voltages and tangential fields decrease the overall flashover voltage [4]. Although the present experiment is configured to a produce a uniform field distribution that is purely tangential to the spacer surface, triple-junction field enhancement can serve to establish normal fields and encourage dispersion of flashover voltages as well as charge migration from the spacer interior to the surface [24].

The most directly comparable study to that described in this paper was performed by Pillai and Hackam [19], who conducted DC flashover studies in vacuum up to atmospheric air for a variety of spacer materials. Although results at these low pressures are not directly applicable to the compressed-air designs currently being pursued, they can provide comparison of trends between the relative flashover voltages of different spacer materials. They found that the permittivity of the spacer material is an important factor influencing flashover voltage due to its effect on the degree of field enhancement at the triple-junction. They also observed that flashover voltage of organic spacers scaled inversely with permittivity; however, the higher permittivity ceramic spacers had significantly higher flashover voltage when compared to organic spacers in vacuum conditions. They surmised that the higher electron impact energy required to initiate secondary emission from the ceramic surface leads to lower overall rates of secondary electron emission in the gas, in turn yielding higher flashover voltages, and they proposed that this effect offsets the increased emission due to field enhancement caused by the higher permittivity of the ceramics at the triple-junction. It is important to note that vulnerability to flashover associated with secondary emission may be mitigated in the present study due to much shorter mean-free-paths. However, higher permittivity may still lead to higher triple-junction field enhancement and more partial discharges, which will likely affect results at higher pressure.

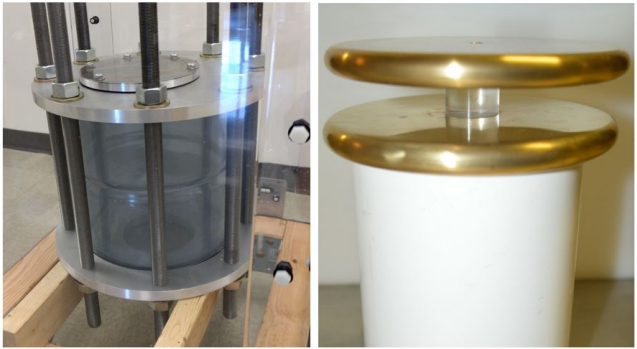

**Figure 1.** (Left:) pressure vessel used to contain the experiment; (right:) the electrode-spacer setup as it would sit in the pressure vessel.

## 3 APPARATUS & METHODS

All experiments discussed herein were conducted in a cylindrical pressure vessel composed of a 30 cm diameter transparent PVC tube and aluminum end plates containing ports for gas and high voltage cabling seen in Figure 1 on the left. Grooves in the end plates are lined with a foam EPDM gasket and are pressed over the ends of the PVC tube with eight 2.5 cm (1"-8) threaded steel rods allowing for gauge pressures up to 414 kPa (60 psig), limited by the pressure rating of the PVC tube.

Within this pressure vessel, a cylindrical spacer is placed between two electrodes, which are supported by another PVC tube, as seen in Figure 1 on the right. The electrodes are 1.27 cm thick, 12.7 cm diameter brass discs with a small threaded hole in the center for connecting high voltage leads. The edges are rounded to provide an evenly distributed electric field in the region between the electrodes. In order to simulate practical laboratory conditions, the electrodes were prepared by sanding only up to 1500 grit sandpaper, a surface roughness that can be expected in day-to-day operations.

The electrodes are positioned 15 cm above the lower end plate by the inner PVC cylinder and the spacer is located at the center of the electrodes. High voltage cable connects the outer faces of the electrodes to high voltage and ground. Pressure is supplied via an NPT connection on the lower end plate and monitored using an analog gauge attached to the upper end plate with a measurement accuracy of +/- 3.5 kPa (0.5 psia). Atmospheric air is compressed and passes through both a moisture filter and a 5 μm particulate filter before entering the pressure vessel. A high voltage power supply (Spellman SL120PN600) is connected to the two electrodes using high voltage cable (Dielectric Sciences 2125) yielding an upper voltage limit of 100 kV and a current measurement accuracy of 10 μA. Cable glands on the aluminum end plates sealed with room-temperature-vulcanizing (RTV) silicone provide both extra electrical insulation and sufficient pressure sealing. To ensure sufficient current supply for formation of an arc, a 3 nF capacitor is connected in parallel with the discharge cell. A simplified circuit diagram of the setup in its operating mode is shown in Figure 2.

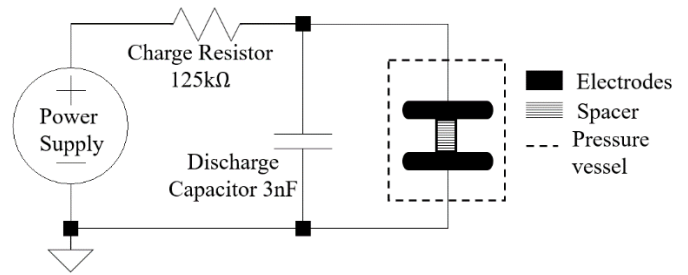

**Figure 2.** Simplified circuit schematic of the experiment. The setup shown in Figure 1 is represented as the electrode, spacer, and pressure vessel here.

In this study, flashover voltage is measured in pressures ranging from 76 kPa (11 psia) to 490 kPa (71 psia) at an ambient temperature of 20° C. These voltages are compared to the Paschen limit and statistical time lag is also measured. Spacer materials tested include G10 ($\epsilon_r \approx 4.7$), acrylic ($\epsilon_r \approx 3.0$), nylon ($\epsilon_r \approx 4.5$), borosilicate ($\epsilon_r \approx 4.5$), and a 100% fill of 3-D printed polyjet resin ($\epsilon_r \approx 1.6$). All spacers are cylindrical with 2.5 cm (1") diameter and lengths of 0.64 cm (0.25"), 1.27 cm (0.5"), and 1.92 cm (0.75"). These values are summarized in Table 1. Acrylic, nylon, and G10 were cut from stock rod and sanded flat on the ends to an overall length tolerance of $\pm 1.5$ mm. Both the borosilicate and 3-D printed spacers were fabricated with the same tolerance or better as the other organic spacers. Since material surfaces could not be prepared in like manner (plastic, glass, fiber-composite), spacer surfaces were left unmodified and the surface roughness of the spacers is that which can be expected from raw materials. This also ensures that the results obtained are applicable to pulsed-power systems fabricated with readily available materials.

**Table 1.** Materials tested.

| Material | Relative Permittivity | Lengths Tested (cm) |
|---|---|---|
| G10 | 4.7 | 0.64, 1.27, 1.92 |
| acrylic | 3.0 | 0.64, 1.27, 1.92 |
| nylon | 4.5 | 0.64, 1.27, 1.92 |
| borosilicate | 4.5 | 0.64, 1.27, 1.92 |
| polyjet resin | 1.6 | 0.64, 1.27, 1.92 |

Setup consisted of placing a spacer between the two electrodes and setting the pressure to a chosen value. The experimental procedure involved first applying 2.5 kV for 35 seconds after which the supply automatically switched off, dropping the potential of the hot electrode to back to ground, and the voltage set-point was then increased by 2.5 kV. Note that the power supply required approximately 5 seconds to reach its target output voltage, yielding an effective voltage exposure time of 30 seconds. Any observed voltage transients during the 30 second exposure time were accompanied by a visible flash and an audible report, which was then recorded as a flashover event. Note that no current transients of magnitude exceeding the power supply's measurement accuracy of $\pm 10$ μA were observed apart from the recorded flashover events. Therefore, any leakage current present in the system is limited to 10 μA or less. Successive data were obtained by starting the voltage set-point at approximately 50% of the voltage at which flashover was first observed, after which the same procedure was followed. Spacers were not cleaned or discharged between flashover events to be more representative of conditions experienced in pulsed-power laboratories so that any effects of residual charge are included in the presented data. Note that the voltage measurement error for all data to ± 1.25 kV. When flashover occurred, a keystroke by the observer recorded the time to breakdown such that the error in the time measurement arises from human reaction time, estimated to be +0.2-1.0 seconds. For each spacer, pressure either was varied from the atmospheric value of 76 kPa (11 psia, standard day atmospheric pressure at Los Alamos altitude of 2200m) to the maximum pressure of 490 kPa (71 psia) or varied from the maximum pressure down to atmospheric. This was done to ensure that the results obtained were consistent for a given pressure and not affected by any possible damage caused to the spacer by the successive flashovers. Five repetitions of flashover initiation were performed for each material, length, and pressure setting.

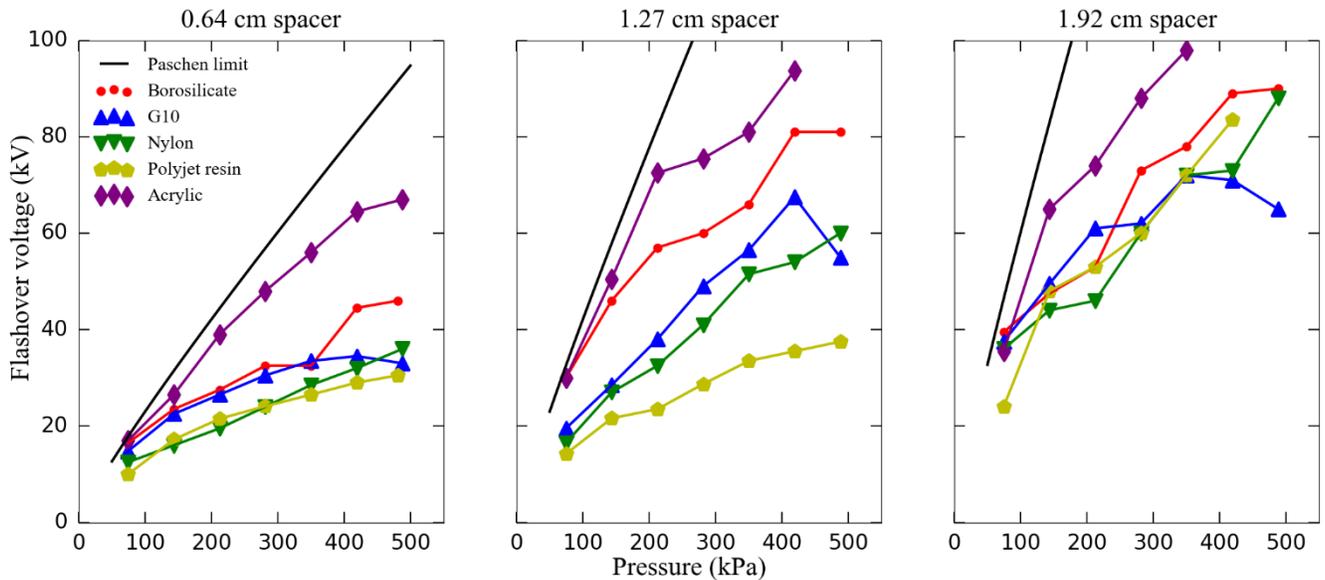

**Figure 3.** Mean flashover voltage plotted against pressure for all tests performed with results for 0.64 cm, 1.27 cm, and 1.92 cm gap lengths plotted from left to right respectively. Each data-point represents the averaged value of five discharges. The Paschen limit is plotted as a solid black line for comparison. The standard deviation of each set of voltage measurements is less than 5 kV.

## 4  RESULTS & DISCUSSION

The mean flashover voltage is plotted for various pressures, materials, and lengths in Figure 3. The solid black line represents the calculated Paschen limit in the absence of a spacer as detailed in Section 2. At each test condition shown in Figures 3 and 5, the standard deviations of the voltage measurements were less than 5kV and no discernible trends were identified. For most spacer materials, the flashover potential increased with pressure but deviated substantially from the Paschen curve at larger distances and pressures. The G10, nylon, and borosilicate have similar relative permittivity and all three materials exhibit similar flashover voltages except in the 1.27 cm (0.5") length case. These results contrast with those obtained by Pillai and Hackam [19] where inorganic spacers were found to have significantly higher flashover voltages compared to organic spacers due to the higher energy requirement for secondary emission from the spacer surface. This indicates that secondary emission from the spacer is not a dominant effect in pressurized systems. However, in the present set of experiments the acrylic spacer with a permittivity of $\epsilon \approx$ 3.0 displayed the highest flashover voltage by far. This demonstrates that field emission at the triple-junction, partial discharge charge deposition, bulk charge migration, or a combination of the three is the primary source of surface charging at higher pressures, leading to a larger analogous ineffective region and further reduced flashover voltages.

The 3-D printed resin exhibited the lowest average flashover voltage despite having the lowest permittivity of all tested materials. Previous work has shown that surface charge accumulation at lower field intensities can be dominated by conductive charge migration [24] as opposed to sources within the gas. Therefore, a material or manufacturing property of the spacer may be facilitating surface charge accumulation through bulk charge migration, leading to lower flashover voltages. However, since the performance of the 3-D printed spacer was found to be unsuitable for our uses these results were not explored further.

A notable exception to the typical trends in Figure 3 is the decrease in flashover voltage of G10 at pressures above 350-450 kPa. Initially the decrease was suspected to be due to surface damage on the G10 resulting from flashover. However, upon visual inspection no damage was observed. To verify that surface damage was not the cause of decreased flashover voltage, another sequence of breakdown testing was performed with increasing *and* decreasing pressure sweeps. The trend was observed to be independent of pressure and flashover history of each G10 spacer, indicating that this effect is not due to damage, rather to another undetermined factor.

Many of the trends in Figure 3 show a distinct change in slope (henceforth referred to as the "knee") after which the breakdown voltage deviates substantially from the Paschen limit. The fact that the knee occurs at different pressures for different materials indicates that it is caused by the spacer and is a material-dependent effect. The knee is present in all curves for the 0.64 cm spacers except for nylon and is present for the acrylic curves for all spacer lengths. Because the slope of the breakdown voltage curves for most materials are roughly similar beyond the knee, it likely that knees exist for other materials at pressures lesser than those investigated here. Previous work investigating corona onset in high pressure air has shown similar trends for the corona's current-voltage relationship [25]; at higher voltages, a slope discontinuity was observed where the slope of the current-voltage trend increased. Higher leakage currents typically lead to lower flashover voltages, which suggests that formation of corona at the triple-junction may be responsible for the knee effect.

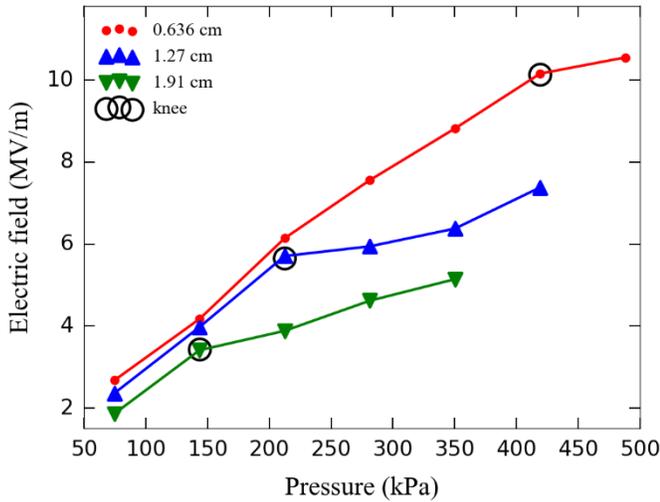

**Figure 4**. Uniform field intensity vs. pressure for all acrylic spacers at flashover. Each data point represents the average of five breakdowns.

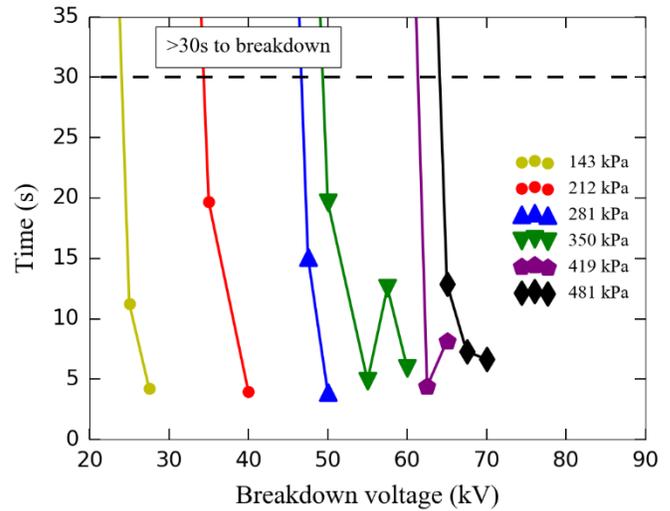

**Figure 5**. The mean time to breakdown after application of voltage plotted against the voltage at which breakdown occurred for the 0.64 cm length acrylic spacer. The standard deviation of each set of voltage measurements is less than 5 kV.

Figure 4 displays the field intensity in the gap at which flashover occurred versus pressure for the acrylic spacers. The field intensity shown is the average field intensity between the electrodes assuming an ideal uniform field distribution across the gap. The knee occurs at lower pressures for larger distances, indicating the decrease in strength is not caused by an interface effect, such as triple-junction field enhancement. Further, the knee for the 0.64 cm lengths of G10 and borosilicate occur at 420 kPa and 280 kPa respectively, despite the materials having nearly the same permittivity. This indicates that the pressure at which the knee occurs is not dependent on spacer permittivity but is still material dependent so that a different material property must influence the pressure associated with the knee location. The ability of the spacer's surface to trap electrons, either due to surface finish or spacer chemistry, may be critical to understanding high-pressure flashover behavior.

Most notably in Figure 4, the average flashover field intensity decreases for larger gap lengths. As distances increase for a given electric field intensity and pressure, it is reasonable to postulate that more partial discharges will occur due to the increased surface area, leading to more surface charge deposition and diminished returns on flashover voltage for increased distances. This effect may also be a determining factor in the knee location.

The statistical time lag to breakdown for DC conditions must also be taken into consideration for design derating. For each pressure and voltage setting in the experiment, the elapsed time prior to initiation of an arc after the application of voltage was recorded. An example, presented in Figure 5, is the average time to breakdown versus voltage at each pressure setting for the 0.64 cm acrylic spacer. Note that other materials exhibited similar statistical time lag trends. These data indicate that the time to breakdown decreases rapidly (approximately exponentially) as gap voltage increases. Note that these data were used to determine sufficient derating of flashover standoffs for designs that require 30 seconds or less at full charge, e.g. charging pulsed-power systems; however, extrapolation of these data is not recommended due to the presence of other effects at longer timescales. For example, previous work has shown that the presence of temperature gradients in the spacer-electrode system due to ohmic heating lead to further reduced flashover voltage [16]. Therefore, for longer timescale applications, testing of statistical time lag in the intended regimes should be conducted.

## 5  CONCLUSIONS

Flashover voltages were measured in compressed air under applied uniform fields for pressures ranging from 76 kPa (11 psia) to 490 kPa (71 psia) and voltages up to 100 kV. Tested spacers included acrylic, borosilicate, G10, nylon, and 3-D printed resin at lengths of 0.64 cm (0.25"), 1.27 cm (0.5"), and 1.92 cm (0.75"). Of the materials tested, acrylic exhibited the highest flashover voltage while 3-D printed resin exhibited the lowest. A discontinuity in slope or "knee" was observed for acrylic, at which the gains in flashover voltage with pressure decreased, deviating from the Paschen limit. These results suggest that the same knee exists at lower pressures for other materials since they do not exhibit slope agreement with the Paschen limit at the pressures investigated here. These results also show that the knee is due to the presence of the spacer and is a material dependent effect. The pressure at which the knee occurs appears to be independent of triple-junction effects as well as material permittivity, indicating that its location depends predominantly on surface properties such as surface finish or spacer chemistry. The electric field intensity required to initiate flashover was also observed to decrease with increasing spacer lengths. It is hypothesized that large surface areas allow for increased surface charging due to partial discharges, thereby decreasing the field strength required for flashover. Data regarding statistical time lag were gathered to inform derating of spacer lengths.

The data obtained in this experiment informs the design of compressed-air insulated, flashover-resistant interfaces, which will facilitate the transition from currently-existing $SF_6$ insulated designs. This information has been applied successfully in designing compressed-air insulated, flashover resistant pulsed-power devices at Los Alamos National Laboratory. However, outstanding questions remain regarding the timescales on which surface charging occurs and how it relates to statistical time lag, what factors are dominant in determining the "knee" location, why the flashover voltage of G10 decreases with increasing pressure above a certain threshold, and whether surface charge accumulation due to partial discharges, triple-junction effects, or charge migration is responsible for reduced flashover voltages at larger distances.

## ACKNOWLEDGMENT

This work is supported by Institute for Critical Technology and Applied Science (ICTAS) at Virginia Tech and the National Nuclear Security Administration of the U.S. Department of Energy under contract 89233218CNA000001. The authors would also like to thank Maximillian Schneider (Virginia Tech), Patrick Crandall (Virginia Tech), Minzhen Du (Virginia Tech), and John Dunn (Los Alamos National Laboratory) for assistance in constructing test materials and operating the experiment. Approved for unlimited release, LA-UR-20-22485.

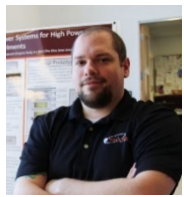

**Ian A. Bean** received his B.S. in aerospace engineering from Virginia Tech in 2017. He is currently a PhD candidate in applied physics in aerospace at Virginia Tech. He is pursuing research in pulsed-power design and plasma physics at Los Alamos National Laboratory as a visiting student.

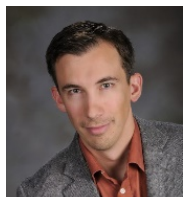

**Colin S. Adams** received his B.S. and M.S. degrees in aerospace engineering from the University of Washington in 2005 and 2009 respectively. Between 2007 and 2011, Adams was an aeronautical engineer in private industry before appointment as a postmaster's researcher in plasma physics at Los Alamos National Laboratory. After resuming his studies at the University of New Mexico, he received his Ph.D. in electrical engineering in 2015. Since then he has been with the Kevin T. Crofton Department of Aerospace and Ocean Engineering at Virginia Tech where he is presently an Assistant Professor, directing the Experimental Plasma and Propulsion Laboratory and the hypersonic wind tunnel. His research interests include plasma physics, fusion energy science, pulsed-power technology development, spacecraft propulsion systems, and hypersonics.

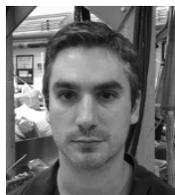

**Thomas Weber** received a B.S. aerospace engineering with a minor in astrophysics from the University of Minnesota in 2005 and went on to earn a M.S. (2007) and Ph.D. (2010) in plasma physics from the department of Aeronautics & Astronautics at the University of Washington. His graduate research included the development of novel high-power plasma thrusters and compact magnetic fusion energy concepts. Upon graduation, he worked as a Guest Scientist and Postdoctoral Associate researching Magnetized Target Fusion and laboratory plasma astrophysics. He is currently a staff scientist in the Physics Division at Los Alamos National Laboratory and leads the Magnetized Shock Experiment (MSX) laboratory conducting research in Magneto-Inertial Fusion (MIF) and radiation science.